\documentstyle[11pt,newpasp,twoside,epsf]{article}
\def\etal{et al.\ }

\def\edcomment#1{\iffalse\marginpar{\raggedright\sl#1\/}\else\relax\fi}
\reversemarginpar

\begin{document}
\title{A CFH12K Survey of Red Giant Stars in the M81 Group}
 \author{Patrick R. Durrell, Megan E. DeCesar, Robin Ciardullo}
\affil{Department of Astronomy \& Astrophysics, Penn State University, University Park, PA USA}
\author{Denise Hurley-Keller, John J. Feldmeier}
\affil{Department of Astronomy, Case Western Reserve University, Cleveland, OH}

\begin{abstract}

We present the preliminary results of a wide-field photometric survey
of individual red giant branch (RGB) and asymptotic giant branch (AGB)
stars in the M81 group, performed with the CFH12K mosaic camera of the
CFHT.  We use deep VI images of 0.65 sq.~deg.~of sky to map out the
two-dimensional distribution of intragroup stars and to search for
stars associated with the many HI tidal tails in the group.  We place
an upper limit on the presence of metal-poor RGB stars in a field
located 50-80 kpc from M81, and derive an `intragroup' fraction of
$<2$\%.  In a field sampling the M81-NGC3077 HI tidal tail, we find
blue stars associated with some of the tidal features, including 2
clumps which we tentatively describe as tidal dwarf candidates.  These
objects are $\sim 1$ kpc in size, and, based on their color-magnitude
diagrams, have formed stars as recently as $\sim30-70$ Myr ago, long
after the group's most recent interactions.

\end{abstract}

Tidal encounters often liberate stars from their parent galaxies
(e.g., Moore \etal 1996).  Intracluster red giants and planetary
nebulae are excellent tracers of such stars and have been observed in
the cluster environments of Virgo and Fornax (Ferguson \etal 1998;
Durrell \etal 2002; Feldmeier \etal 2003; Arnaboldi \etal 2003), where
high-speed encounters are important.  However, little is known about
their presence in smaller groups: do `intragroup' stars exist?  The
M81 system is the nearest group of galaxies to have undergone recent
interactions ($\sim 220-280$ Myr ago; Yun 1999), and the group is
filled with tidally-produced HI (Yun
\etal 1994).  As such, it provides an excellent laboratory to search
for stars extracted from previous encounters.


We have performed a deep $VI$ imaging survey (using the CFH12K camera
at the CFHT; field size 0.33 sq. degrees) to resolve the brightest
stars in the M81 group.  Two group fields have been imaged, with a
third field (located $10\deg$ away) taken to quantify the sizeable
contamination from background galaxies.  Typical limiting magnitudes
of our survey are $V \sim 25.5$, $I\sim 24.5$.

Since our Field 2 data (located $50-80$ kpc from M81) are only just deep
enough to reach the RGB tip, we can only place a
rough upper limit on the presence of intragroup stars.  Using the $VI$ color
magnitude diagram, number counts for all {\it unresolved} objects in
the 5 deepest chips of the CFH12K array, and a statistical subtraction
of a background CMD, we find an excess of
$1.0\pm 0.4$ RGB stars arcmin$^{-1}$.  This extrapolates to a surface
brightness of $\mu_V \sim 31$.  If this luminosity density is constant
over the $\sim 7$ sq.~deg. of the group's core, then the total intragroup 
$V$-band luminosity is $< 6 \times 10^8 L_{\sun}$,
or $< 2$\% of the group's total luminosity.  Note that this 
estimate does not include the likely presence of M81 halo stars nor the 
contribution of a metal-rich population.

\begin{figure}[h]
\plottwo{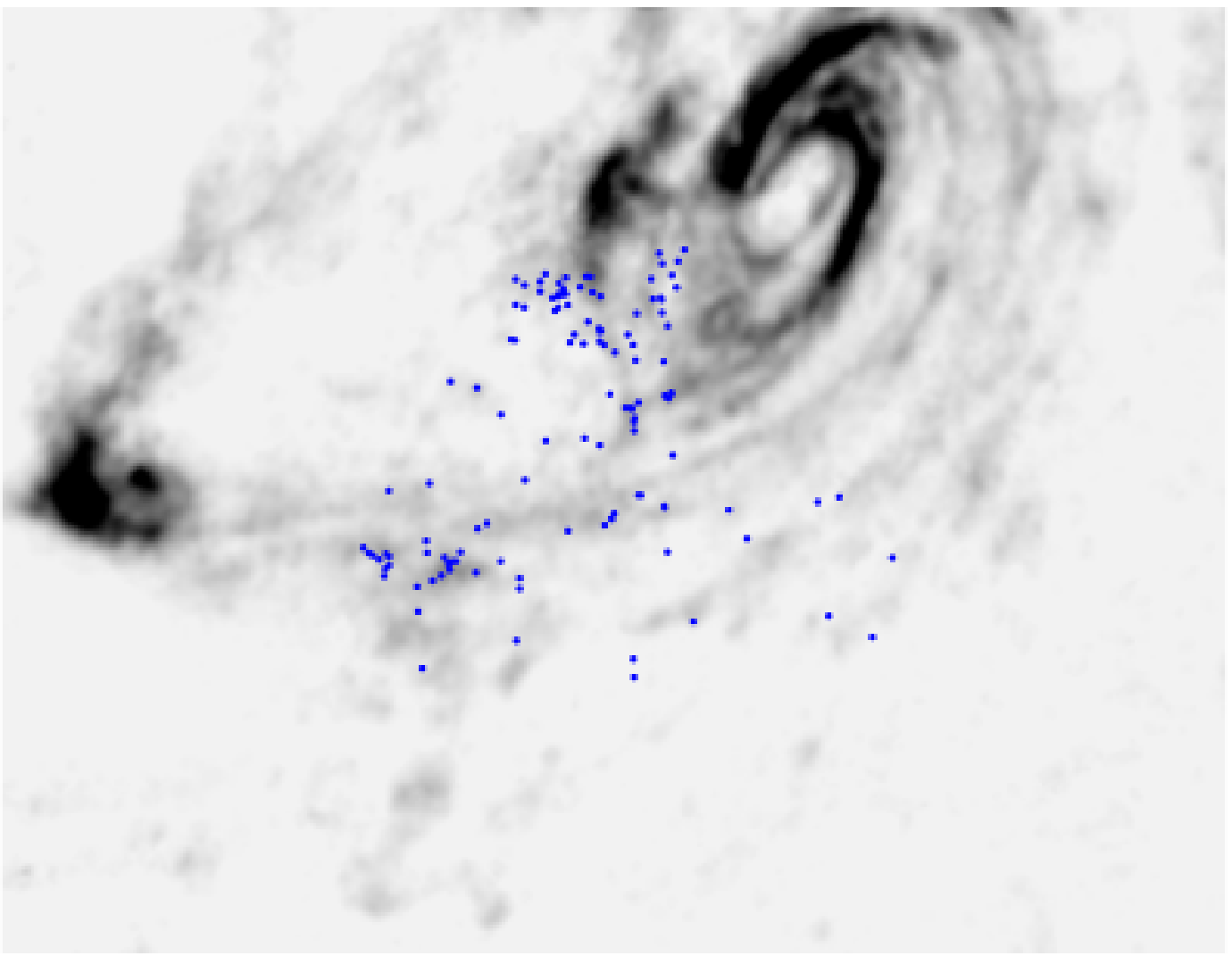}{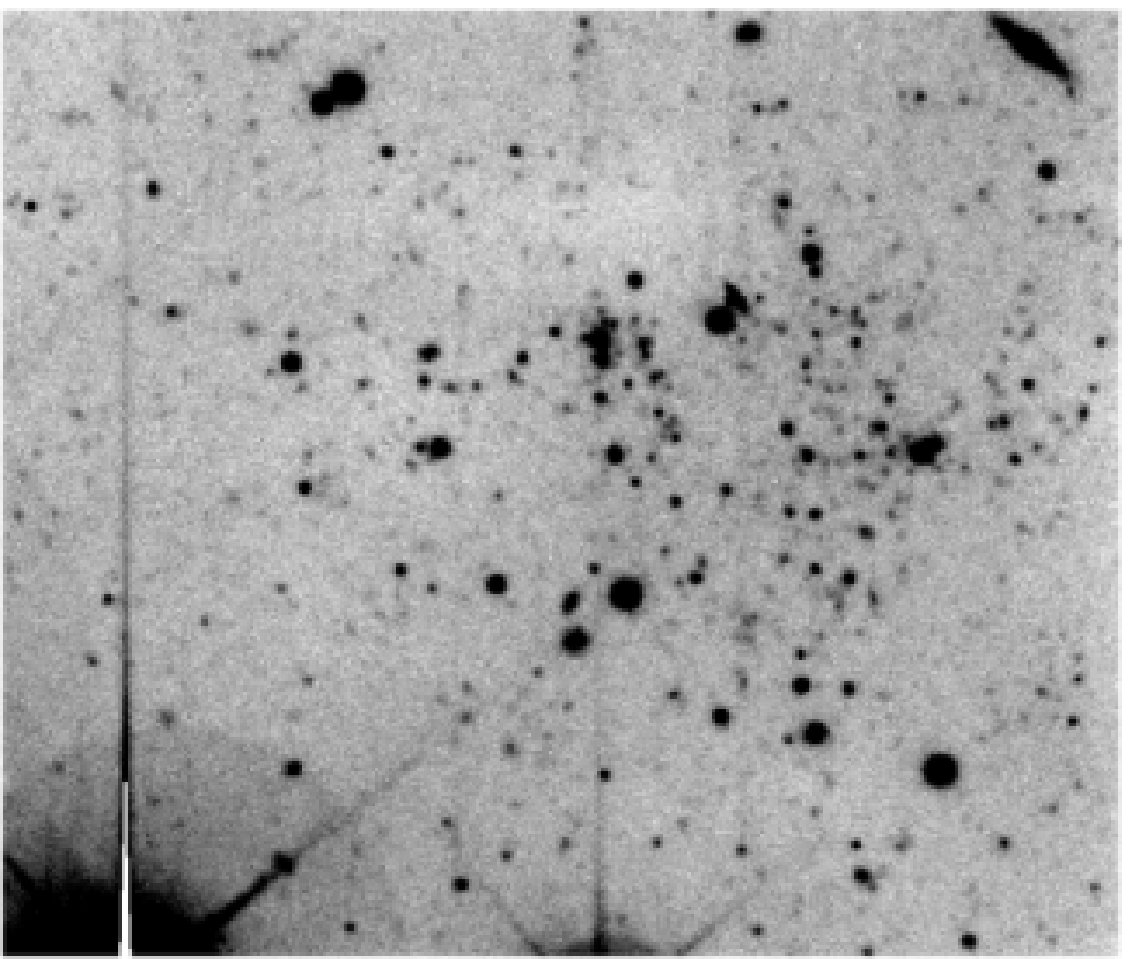}
\caption{Left: Locations of blue stars in our Field 1 CMD, 
superimposed on the HI map of Yun \etal (1994).  Contamination from 
foreground stars and background galaxies is $\sim 20$\%.   
Right: $V$ image of the region in the HI South Tidal Arm that contains the 
blue star clump.}
\end{figure}


Our data also indicate the presence of blue ($V$$-$$I$$<$$0.4$) stars
associated with some of the group's HI tidal streams.  As shown
in Fig. 1, there is a clump of blue objects located in the densest HI
feature of the Southern Tidal Arm.  This feature, which is considered an
`HI satellite' by Yun \etal (1994), is about $\sim$$1\arcmin$ (1
kpc) across.  Isochrones matched to the object's CMD, and that of an adjacent
clump, suggest that star formation occurred $30-70$ Myr ago, well after
the tidal encounters which created the HI arm.  These star-forming
clumps are tidal dwarf candidates.

\acknowledgments{We would like to thank Chris Mihos and George Jacoby 
for related discussions, and Min Yun for providing us with 
his HI map.}

\end{document}